\documentstyle[12pt]{article}
\newcommand{\ra}{\rightarrow}

\begin{document}
\begin{center}{\bf Estimates of flavoured scalar production in B - decays}
\end{center}
\vspace{0.5cm}
\begin{center} {\bf Victor Chernyak}\end{center}
\begin{center}Budker Institute of Nuclear Physics,\, Novosibirsk, 630090,
Russia\end{center}
\vspace{1cm}
\begin{center}Abstract\end{center}

\begin{center} Estimates are presented for the branching ratios of several \\
two-particle B-meson decays into flavoured scalar mesons.
\end{center}
\vspace{1cm}

 It seems that there are no estimates of B-meson decays into
scalar mesons. The purpose of this short note is to present such 
estimates. As will be shown, some two-particle decays of B-mesons into
scalar mesons have sufficiently large branchings to be of current interest.\\
 
{\bf 1.} Having in mind $B^{-}\ra {\bar K}_{o}(1430)\pi^{-}$ and 
$B^{-}\ra {\bar K}\pi^{-}$ decays, the main contributions in the 
factorization approximation which looks reliable for these decays, and in 
the standard notation, come from two terms in the effective Hamiltonian:
$$
H_{eff}=\frac{G_F}{\sqrt 2}(-V_{tb}V^{*}_{ts})\left \{ a_4\,[\,{\bar s}
\gamma_{\mu}(1+\gamma_{5})d\,]\otimes [\,{\bar d}\gamma_{\mu}(1+
\gamma_{5})b\,]\,-\right.
$$
$$
\left.  -\,2\,a_6\,[\,{\bar s}(1-\gamma_5)d\,]\otimes [\,{\bar d}
(1+\gamma_5)b\,]\right \}.  \hspace{3cm} (1)
$$

The effective coefficients $a_i$ in (1) can be expressed either explicitly
through the original coefficients $C_i(\mu)$ of the effective Hamiltonian
\cite{Bur}, plus perturbative one loop corrections: $C_i(\mu)\ra C_i^{eff}$;
this cancels the main scale dependence of $C_i(\mu)$, see e.g. \cite{Ali}
and references therein. Or they can be calculated in "the QCD improved
factorization approximation" \cite{Ben}, which is not much different. 

In what follows we will not need their explicit form, but only the typical 
value of the ratio $a_6/a_4$.
\footnote{
The coefficients $a_i$ change, of course, when going from $K$ to $K_o(1430)$,
as they depend on the form of the meson wave functions. In our estimates 
below we suppose these differences can be safely neglected for $a_4,\,a_6$ 
and $a_1$. Using the explicit expressions for $a_i$ from \cite{Ben}, which 
include $O(\alpha_s)$ corrections, and the model wave functions of $K$ and 
$K_o(1430)$, we have estimated that the differences in the values of $a_{1,4,
6}$ for $K$ and $K_o(1430)$ are indeed reasonably small. 
}

The matrix elements are defined as:
$$
\langle {\bar K}(q)|{\bar s}\gamma_{\mu}\gamma_5 d|0\rangle =-if_k\,q_{\mu},
\quad \langle {\bar K}(q)|{\bar s}\,i\gamma_5 d|0\rangle=f_k\,\mu_k,\quad 
\mu_k=M_k^2/{\bar m}_s,
$$
$$
\langle {\bar K}_o(1430)(q)|{\bar s}\gamma_{\mu} d|0\rangle =f_o\,q_{\mu}, 
\quad \langle {\bar K}_o(1430)(q)|{\bar s}\,d|0\rangle=f_o\,\mu_o,\quad 
\mu_o=M_{o}^2/{\bar m}_s,
$$
$$
\langle \pi^-(p_2)|\,{\bar d}\gamma_{\mu}b\,|B^-(p_1)\rangle =
(p_1+p_2 )_{\mu}\,f_{+}(q^2)+q_{\mu}\,f_{-}(q^2)\,,
$$
$$
M_b\,\langle \pi^-(p_2)|{\bar d}\,b|B^-(p_1)\rangle 
\simeq (M_B^2-q^2)\,f_{+}^{B\pi}(q^2)\simeq M_B^2\,f_{+}(0). \hspace{3cm} (2)
$$ 

Therefore, the decay amplitudes look as:
$$ T(B^-\ra {\bar K}\pi^-)\simeq i\,\frac{G_F}{\sqrt 2}V_{tb}V^{*}_{ts}
\,a_4 f_k M_B^2 f_{+}^{B\pi}(0)\left (1+\frac{2 M_k^2}
{{\bar m_s}M_b}\,\frac{a_6}{a_4}\right ),
$$
$$
T(B^-\ra {\bar K}_o\pi^-)\simeq \frac{G_F}{\sqrt 2}V_{tb}V^{*}_{ts}(-a_4)
f_o M_B^2 f_{+}^{B\pi}(0)\left (1-\frac{2 M_{o}^2}{{\bar m_s}M_b}
\,\frac{a_6}{a_4}\right ).\quad\quad (3)
$$

Now, about parameters entering (3). To avoid main uncertainties it is 
reasonable to take
the ratio $Br(K_o\pi)/Br(K\pi)$. So, it will be sufficient to know 
($f_k\simeq 160\,MeV)$:
$$
\frac{a_6}{a_4}\simeq 1.2-1.3,\quad {\bar m}_s\simeq m_s(\mu=2.5\,GeV)\simeq
110\,MeV,\quad  M_b^{(eff)}\simeq 4.8\,GeV\,. \,\,(4)
$$

The main new parameter is the coupling $f_o$. It can be estimated from
the form factor ($\Delta M^2\equiv M_k^2-M_{\pi}^2\simeq 0.22\,GeV^2)$:
$$
\langle K^-(p_2)|{\bar s}\gamma_{\mu}d|\pi^-(p_1)\rangle=
\left (p_1+p_2-(\Delta M^2/q^2)q\right )_{\mu}F_{+}(q^2)+(\Delta M^2/q^2)
q_{\mu}F_o(q^2)\,,
$$
$$
m_s \langle K^-|{\bar s}\,d|\pi^-\rangle \equiv d(q^2)=\Delta M^2
F_o(q^2),\, F_o(q^2=0)\simeq 1\,.  \quad\quad (5)
$$

Saturating the dispersion relation for $d(q^2)$ by two lowest resonances,
$K_o(1430)$ and $K_o(1950)$, one obtains:
$$
0.22\,GeV^2\simeq \left ( f_o\,g_o+f^{\prime}\,g^{\prime}\right )\,,
$$
where $f_i$ are couplings of resonances with the scalar current (see eq.(2)
above, the coupling $f^{\prime}$ of $K_o(1950)$ with the current $m_s({
\bar s}d)$ is defined in the same way as those of $K_o(1430);\, f_i=O(m_s)$
at $m_s\ra 0)$, and $g_i$ are 
their couplings to the $(K^{-}\pi^{+})$-pair. These last can be found 
from their known decays to $K\pi: g_o\simeq 3.8\,GeV,\,g^{\prime}\simeq 2.7
\,GeV$. Besides, there are estimates \cite{Jam} of the ratio of couplings 
$f^{\prime}/f_o$ which look as:
$$
\frac{f^{\prime}}{f_o}= -\gamma\,\frac{M_o^2}{M{^\prime}^2}= 
-0.5\,\gamma\,, \quad \gamma=(0.5\pm 0.3)\,.
$$

Therefore, one obtains: 
$$
f_o=(70\pm 10)\,MeV.\hspace{4cm}(6)
$$

Collecting all the above given numbers, we have:
$$
Br\,\frac{(B^-\ra {\bar K}_o(1430)\pi^-)}{(B^-\ra {\bar K}\pi^-)}\simeq
14\,\frac{f_o^2}{f_k^2}\simeq 2.7
$$
(for central values of parameters in (4) and (6)). 

So, if $Br(B^-\ra {\bar K}\pi^-)\simeq 16\cdot 10^{-6}$, then $Br(B^-\ra 
{\bar K}_o(1430)\pi^-)$ will be $\simeq 43\cdot 10^{-6}$.

It is interesting not only that $Br(B^-\ra {\bar K}_o(1430)\pi^-)$ 
is large by itself, but that it receives the dominant contribution from the
term $\sim a_6$ which is a power correction, $O(\Lambda_{QCD}/M_b)$, in
the formal limit $M_b\ra \infty$.\\

{\bf 2.} Let us consider the decay ${\bar B}^o \ra a_o^{+}(1450)K^-$. 
The corresponding form factor $F_{+}^{Ba}$ is defined as:
$$
\langle a_o^{+}(p_2)|{\bar u}\gamma_{\lambda}\gamma_{5}b|{\bar B}^o(p_1) 
\rangle = -i\left [(p_1+p_2)_{\lambda}
F_{+}^{Ba}+q_{\lambda}F_{-}^{Ba}\right ]\,,
$$
$$
M_b\langle a_o^{+}|{\bar u}\gamma_5b|{\bar B}^o\rangle\simeq i\,M_B^2
F_{+}^{Ba}(0)\,. \hspace{3cm}(7)
$$

Such form factors, at not too large $q^2$, can be found by 
the method proposed in \cite{Ch} (which is known now as "the light-cone
sum rules"). One considers the correlator:
$$
K=i\int dx e^{iqx}\langle a_o^{+}(1450)(p)|T\{{\bar u}(x)\gamma_5 b(x)\,,
{\bar b}(0) \gamma_5 d(0)|0\rangle\,,
$$
and proceeding as in \cite{Ch} obtains the sum rule ($\Delta=(1-M_b^2/S_o)
\simeq 0.3)$:
$$
F_{+}^{Ba}(0)\simeq \frac{M_b^3\, {\bar \lambda}^2}{f_B M_B^4}\int_0^{\Delta}
dx\frac{\phi_s(x)}{1-x}\exp\left \{\frac{M_B^2-\frac{M_b^2}{1-x}}{M^2}
\right \}\,,\quad (8)
$$
where the wave function $\phi_s(x)$ of $a_o^{+}(1450)$ is defined as: 
\footnote{
The leading twist wave function of $a_o$ also contributes to the sum rules.
Estimates show that this contribution is positive and small. We neglect it.
}
$$
\langle a_o^{+}(1450)(p)|{\bar u}(0)\,d(z)|0\rangle ={\lambda^2}\int_0^
1 dx e^{ixpz}\phi_s(x),\quad \int_0^1 dx \phi_s(x)=1.
$$
The coupling ${\bar \lambda}^2=\lambda^2(\mu\simeq 1.5\,GeV)$ 
is related by SU(3) to the matrix element 
$\langle K_o^*(1430)(q)|{\bar s}\,d|0\rangle$ in (2), and so: 
${\bar \lambda}^2\simeq 1.15\,GeV^2.$ For other quantities entering 
(8) we use: $\phi_s(x)\simeq \phi_s^{asy}(x)= 1,\,f_B\simeq f_{\pi}
\simeq 130\,MeV$. One obtains then from (8):
$$
F_{+}^{Ba}(0)\simeq 0.46\,\,\,. \hspace{4cm} (9)
$$
Somewhat surprisingly, this transition form factor turns out to be $\simeq
1.5$ times larger than the corresponding $B\ra \pi$ form factor: $f_{+}^
{B\pi}(0)\simeq 0.30$. Finally, this is due to strong coupling of scalar 
mesons to the scalar current.

Proceeding now in the same way as above, one obtains the decay amplitude:
$$
T({\bar B}^o\ra a_o^{+}(1450)K^-)\simeq \frac{G_F}{\sqrt 2}V_{tb}V^{*}_{ts}
\,a_4 f_k\,M_B^2\,F_{+}^{Ba}(0)\left [\,1-\frac{a_6}{a_4}\,
\frac{2M_k^2}{{\bar m}_s M_b}\right ]. \,\,\, (10)
$$
The two terms in square brackets in (10) nearly cancel each other. So, we
conclude that $Br ({\bar B}^o\ra a_o^{+}(1450)K^-)$ is very small, in spite
of the large form factor.\\

{\bf 3.} Let us consider now the decay ${\bar B}^o\ra a_o^{+}(1450)
\pi^-$. Proceeding as before, one obtains the decay amplitude (it follows 
from the above that the penguin contribution is negligible):
$$
T({\bar B}^o\ra a_o^{+}(1450)\pi^-) \simeq  \frac{G_F}{\sqrt 2}V_{ub}V^{*}_
{ud}\,(-a_1)\,f_\pi M_B^2 F_{+}^{Ba}(0)\,.\quad (11)
$$
One has then from (11) and (9): 
\footnote{
$Br ({\bar B}^o\ra a_o^{-}\pi^{+})$
is highly suppressed, and so this mode is selftagging.
}
$$
Br ({\bar B}^o\ra a_o^{+}(1450)\pi^-) \simeq 20\cdot 10^{-6};\,\,\,
Br ({\bar B}^o\ra a_o^{+}(1450)\rho^-) \simeq 38\cdot 10^{-6}\,.
$$
It is seen that these branchings are sufficiently large to be observable.\\

{\bf 4.} Finally, let us consider production of two scalar mesons, $ B^-
\ra {\bar K}_o(1430)a_o^{-}(1450).$ Proceeding as before, one obtains from 
(1) the decay amplitude:
$$
T( B^-\ra  {\bar K}_o(1430)\,a_o^{-}(1450))\simeq i\,\frac{G_F}{\sqrt 2}
V_{tb}V^{*}_{ts}\,a_4 f_o M_B^2 F_{+}^{Ba}(0)\left (1+\frac{2 M_{o}^2}
{{\bar m_s}M_b}\,\frac{a_6}{a_4}\right ).
$$
Normalizing by $B^-\ra {\bar K}\pi^-$ as before, and using (6),\,(9) one
obtains:
\footnote{
The role of the tree $b\ra u\,({\bar u}s)$ transition is very small here in
comparison with the large contribution  $\sim a_6$.
}\\
$$
Br\,(B^-\ra {\bar K}_o(1430)a_o^{-}(1450))\simeq Br\,({\bar B}^o\ra 
 K_o^-(1430)a_o^{+}(1450))\simeq
$$
$$
\simeq 8.8\,\, Br\,(B^-\ra {\bar K}\pi^-)\,.
$$
Therefore, for $Br(B^-\ra {\bar K}^o\pi^-)\,\simeq 16\cdot 10^{-6}$, these
branchings (as well as their charge conjugates)
will be $\simeq 140\cdot 10^{-6}$. It is seen that, in a sense, they are 
very large.\\

{\bf 5.} We do not consider here the neutral decay modes like, for instance, 
$ B^o\ra J/\Psi  K_o(1430)$. Because the main factorizable
contributions cancel each other here to large extent, the non-factorizable
contributions become of great importance, and these are under poor control 
for such decays at present. One can expect only that, because the transition
form factor $B\ra K_o(1430)$ is considerably larger than those of $B\ra K$, 
this mode can hardly be much smaller than $B\ra J/\Psi K$.

We did not consider also flavourless scalars $f_o(980),\,f_o(1370),\,
f_o(1500)$, etc. There are two reasons for this. First, their nature and
quark-gluon composition are not well understood at present and, it seems, 
are complicated. The main reason, however, is that we expect their 
production in the $B\ra f_i K$ decays can be highly enhanced 
by the same mechanism which enhances $B\ra \eta^{\prime}K$, and there is 
no clear understanding of this mechanism up to now.

\begin{center}{\bf Acknowlegements} \end{center}
I am indebted to A.E. Bondar who insisted on writing this note. I also thank
him for very useful discussions.

\end{document}